\documentclass[11pt,a4paper]{article}

\pdfoutput=1

\usepackage{fullpage}

\usepackage{amsmath}
\usepackage{graphicx}
\usepackage[colorlinks,citecolor=red,urlcolor=red]{hyperref}

\newcommand{\myAuthor}[1]{{\sc #1}}
\newcommand{\ead}[1]{\href{mailto:#1}{\tt #1}}
\newcommand{\figName}[1]{#1}

\newcommand{\bmath}[1]{\mbox{\boldmath$#1$}}
\newcommand\vek[1]{\mathbf{#1}}
\newcommand{\tenss}[1]{\bmath{#1}} 
\newcommand\de[1]{\,{\mathrm d}#1} 
\newcommand{\avgs}[1]{\left\langle #1 \right\rangle} 
\newcommand{\evek}[1]{\{\mathsf{#1}\}}
\newcommand{\emtrx}[1]{\left[\mathsf{#1}\right]}
\newcommand{\trn}{{\sf ^T}}

\newcommand{\on}{\mbox{ on }}
\newcommand{\ins}{\mbox{ in }}

\title{Evaluation of Effective Thermal Conductivities of Porous Textile Composites}

\author{%
\myAuthor{Blanka Tomkov\'{a}}$^1$, 
\myAuthor{Michal \v{S}ejnoha}$^{2,3}$, 
\myAuthor{Jan Nov\'{a}k}$^{2,3}$ and
\myAuthor{Jan Zeman}$^2$\thanks{%
Corresponding author, Tel.:~+420-2-2435-4482; fax~+420-2-2431-0775, \emph{E-mail addresses}:
\ead{tomkova@tulib.cz}, \ead{sejnom@fsv.cvut.cz}, \ead{novakj@cml.fsv.cvut.cz}, \ead{zemanj@cml.fsv.cvut.cz}}
}

\date{%
$^1$Department of Textile Materials, Technical University in Liberec\\
H\'{a}lkova 6, 461 17 Liberec 1, Czech Republic
\\[2mm]
$^2$Department of Mechanics, Faculty of Civil Engineering\\
Czech Technical University in Prague\\
Th\' akurova 7, 166 29 Prague 6, Czech Republic
\\[2mm]
$^3$Centre for Integrated Design of Advances Structures\\
Th\' akurova 7, 166 29 Prague 6, Czech Republic
}

\begin{document}

\maketitle

\begin{abstract}
An uncoupled multi-scale homogenization approach is used to estimate
the effective thermal conductivities of plain weave C/C composites
with a high degree of porosity. The geometrical complexity of the
material system on individual scales is taken into account through the
construction of a suitable representative volume element (RVE), a
periodic unit cell, exploiting the information provided by the image
analysis of a real composite system on every scale. Two different
solution procedures are examined. The first one draws on the classical
first order homogenization technique assuming steady state conditions
and periodic distribution of the fluctuation part of the temperature
field. The second approach is concerned with the solution of a
transient flow problem. Although more complex, the latter approach
allows for a detailed simulation of heat transfer in the porous
system. Effective thermal conductivities of the laminate derived from
both approaches through a consistent homogenization on individual
scales are then compared with those obtained experimentally. A
reasonably close agreement between individual results then promotes
the use of the proposed multi-scale computational approach combined
with the image analysis of real material systems.
\end{abstract}


\vfill

\par\noindent 
Accepted in \emph{International Journal for Multiscale Computational
  Engineering}

\clearpage

\section{Introduction}\label{sec:intro}
Carbon-carbon (C/C) plain weave fabric composites belong to an
important class of high-temperature material systems. An exceptional
thermal stability together with high resistance to thermal shocks or
fracture due to rapid and strong changes in temperature have made
these materials almost indispensable in a variety of engineering
spheres including aeronautics, space and automobile industry.
Applications include components in spacecraft protective shields, wing
leading edges or parts of jet aircrafts turbine engines.

\begin{figure}
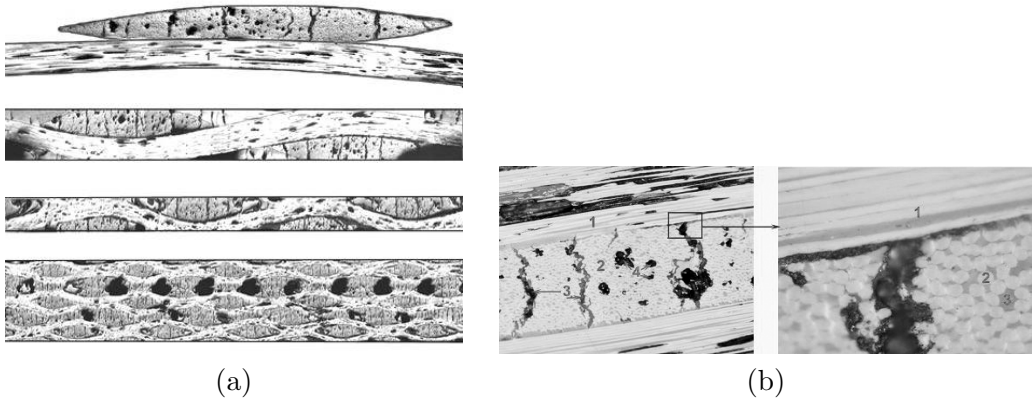

\begin{center}
\begin{tabular}{c@{\hspace{5mm}}c}
\includegraphics[width=6cm]{\figName{figure1a}} &
\includegraphics[width=7cm]{\figName{figure1b}}\\
(a)&(b)
\end{tabular}
\caption{Color images of a real composite system: (a) Scheme of multiscale structural model 
(from top - transverse and longitudinal view of fiber tow composite, composite unit cell, 
composite lamina, composite plate), (b) Carbon tow microstructure showing major pores 
and transverse cracks.}
\label{F1}
\end{center}
\end{figure}

While their appealing thermal properties such as low coefficients of
thermal expansion and high thermal conductivities are known, their
prediction from the properties supplied by the manufacturer for
individual constituents is far from being trivial since these systems
are generally highly complicated. Apart from a characteristic
three-dimensional (3D) structure of textile composites the geometrical
complexity is further enhanced by the presence of various
imperfections in woven path developed during the manufacturing
process. A route for incorporating at least the most severe
imperfections in predictions of the mechanical properties of these
systems has been outlined in~\cite{Zeman:2004:RC} in the context of
statistically equivalent periodic unit cell. Although properly
accounting for three-dimensional effects the resulting representative
volume element still suffers from the absence of the porous phase,
which in real systems, Fig.~\ref{F1}, may exceed 30$\%$ of the overall
volume. As suggested in~\cite{Palan:2002}, neglecting the material
porosity may severely overestimate the resulting thermal properties of
textile composites. In this regard, the procedure introduced in the
current paper extends the previous studies, as they either neglect the
porous phase~\cite{Dasgupta:1996:3DM,Woo:2004:TCCP} or analyze the
porosity effects on the level of fibers and tows
only~\cite{DelPuglia:2004:MDT}.

The porosity of C/C composites directly arises as a result of the
manufacturing process characterized by thermal decomposition and
transformation of an initial polymeric precursor into the carbon
matrix through several steps of carbonization, re-impregnation and
final graphitization. As evident from Fig.~\ref{F1} the major
contribution to the porosity is due to crimp voids and delamination
cracks, which are usually classified as inter-tow voids
(Fig.~\ref{F1}(a)), as well as due to intra-tow voids represented by
pores and transverse cracks developed within the fiber tow composite
(Fig.~\ref{F1}(b)). For more details the reader is referred
to~\cite{Tomkova:2006} and references therein.

While simplifying averaging schemes may provide rational
micromechanics models for the prediction of effective thermal
conductivities~\cite{Benveniste:JAP:90} at the level of the fiber tow
composite in Fig.~\ref{F1}(b), the complex mesoscopic structure of
woven fabric plotted in Fig.~\ref{F1}(a) calls for considerably more
accurate treatment of the actual geometry as already demonstrated
in~\cite{Zeman:2004:RC,Zeman:MSMSE:2007}. In such a case, the image
analysis combined with a reliable morphological description of the
underlying composite structure then provide a general tool for the
determination of what has been termed the statistically equivalent
periodic unit cell (SEPUC) introduced by the authors in their previous
works on random and imperfect
composites~\cite{Zeman:2001:EPG,Zeman:2004:RC,Zeman:MSMSE:2007}. Unlike
classical averaging schemes, information on the local fields in
representative volume elements formulated on the bases of SEPUCs is
derived through a detailed numerical analysis which typically employs
the finite element method (FEM). A special treatment of boundary
conditions is then needed to establish a link with the first-order
homogenization scheme which develops upon the assumption of
homogeneous effective (macroscopic) fields~\cite{Ozdemir:IJNME:2008}.

Existence of macroscopically uniform fields (strains or stresses in
the case of mechanical problem or temperature gradient and heat flux
in the case of heat conduction problem) readily allows for splitting
the local fields into macroscopic and fluctuation parts which in view
of the solution of heat conduction problem reads
\begin{equation}
\theta(\tenss{x}) = \tenss{H}\cdot\tenss{x} +\theta^{*}(\tenss{x})\quad{\rm or}\quad
\theta(\tenss{x}) = H_ix_i + \theta^{*}(\tenss{x}),\label{eq:Tloc-1}
\end{equation}
where $\tenss{H}$ represents the macroscopically uniform temperature
gradient vector and $\theta^{*}(\tenss{x})$ is the fluctuation part of
the local temperature $\theta(\tenss{x})$. Following,
e.g.~\cite{Michel:1999:EPC,Sejnoha:Habil} the solution of
Eq.~\eqref{eq:Tloc-1} then turns into the search for $\theta^{*}$ in
terms of the applied macroscopic uniform temperature gradient
$\tenss{H}$ or the macroscopic uniform heat flux $\tenss{Q}$.
Consistency between the macroscopic (homogenized) quantities and
volume averages of the corresponding local fields then requires either
setting the boundary values of $\theta^{*}$ equal to zero or
subjecting the fluctuation part of temperature field to periodic
constrains. While both types of boundary conditions are equally
applicable ensuring the macroscopic heat flux being equal to the
volume averaged microscopic (local) heat flux, the latter conditions
will be employed in Section~\ref{sec:theory} when developing the
framework for heat conduction problems as they were shown to provide
the best approximation for a fixed RVE size in a purely mechanical
analysis, see e.g.~\cite{Sluis:MM:2000,Terada:2000:MSC}. The need for
periodic boundary conditions also arises when departing from
asymptotic homogenization, see e.g.~\cite{Fish:1999:CDM}.

It is also interesting to point out that Eq.~\eqref{eq:Tloc-1}, when
applied to either constrains on $\theta^{*}$, is consistent with so
called affine boundary conditions represented in our particular case
by homogeneous temperature or flux applied on the outside boundary
$\Gamma$ of the RVE as
\begin{eqnarray}
\overline{\theta}(\tenss{x}) &=& \tenss{H}\cdot\tenss{x},\hspace{1cm}\overline{q}_n(\tenss{x})=\tenss{Q}\cdot\tenss{n},\hspace{1cm}\on\Gamma,\nonumber \\ 
\overline{\theta}(\tenss{x}) &=& H_ix_i,\hspace{1.25cm}\overline{q}_n(\tenss{x})={Q}_i{n}_i,\hspace{1.15cm}\on\Gamma,\label{eq:THbc-1}
\end{eqnarray}
where $\tenss{n}$ is the outer normal to $\Gamma$. Although the
solution of a steady state heat conduction problem driven by
prescribed macroscopic temperature gradient (or boundary temperatures
consistent with Eq.~\eqref{eq:THbc-1}$_1$), especially if the phase
thermal conductivities are temperature independent, provides directly
the effective conductivity matrix, it appears useful, particularly for
more complex geometries as those in Fig.~\ref{F1}(a), to run the time
dependent transient heat conduction problem which allows for
extracting considerably more information regarding thermal behavior of
the composite material. Quite often, as will also be the case in this
study, a commercial code, which does not allow for direct introduction
of periodic boundary conditions, is used. In such a case, the boundary
conditions given by Eq.~\eqref{eq:THbc-1} prove particularly useful as
they naturally ensure that the volume average of microscopic (local)
fields (temperature gradient or heat flux) are equal to their
macroscopic (prescribed) counterparts $\tenss{H}$ and $\tenss{Q}$ as
evident from Eqs.~\eqref{eq:volume_h} and~\eqref{eq:volume_q}.

To construct a certain periodic unit representing an entire laminate
with all relevant geometrical details (distribution fibers within the
fiber tow composite, waviness of fiber tow path, porosity, etc.) 
might, however, prove rather impractical particularly from the
computational point of view. Instead a so called uncoupled multi-scale
approach~\cite{Fish:00:IJCSE}, still at the forefront of material
science interest, seems rather attractive allowing us to address the
material complexity separately at different levels. Three particular
levels of interest can be identified for the textile composite under
consideration. Henceforth, they will be referred to as micro-scale
(the level of individual fibers within a fiber tow, Fig.~\ref{F1}(b)),
meso-scale (the level of a fiber tow composite,
Fig.~\ref{F1}(a)$_{1-3}$ from top to the bottom) and macro-scale (the
level of a laminated plate, Fig.~\ref{F1}(a)$_{4}$), respectively.  To
estimate a response of a such complex structure it appears reasonable
to perform a sequence of uncoupled analyses corresponding to
individual scales. For this approach to be successfully utilized it is
then viable to establish a link between individual scales. Here, the
concept of scale separation plays a crucial role in the sense that the
three analyses can be carried out independently such that output from
one is used as an input to the other in terms of volume averages of
the local fields while taking into account the boundary conditions
mentioned in the above paragraphs. Such an approach is also adopted in
the present study leading to a consistent search for the effective
(macroscopic) thermal conductivities of a plain weave highly porous
fabric composite.

The paper is organized as follows. Following the introductory part our
attention is paid in Section~\ref{sec:RVE} to the formulation of
various unit cells associated with individual scales. Theoretical
formulation of the homogenization procedure for a steady state heat
conduction problem is outlined in
Section~\ref{sec:theory}. Section~\ref{sec:res} then illustrates the
efficiency and reliability of the applied multi-scale analysis by
comparing the numerical results with those derived experimentally.
The essential findings are finally summarized in
Section~\ref{sec:con}.

\section{Image analysis and construction of the geometrical model}\label{sec:RVE}
It has been demonstrated in our previous work, see
e.g.~\cite{Zeman:2001:EPG,Zeman:2004:RC,Sejnoha:MCE:2004,Sejnoha:2007:MSH}
that image analysis of real, rather then artificial, material systems
plays an essential role in the derivation of a reliable and accurate
computational model. This issue is revisited here for the case of
woven fabric C/C laminate with particular relation to the adopted
uncoupled multi-scale solution strategy.

\begin{table}[ht]
\caption{Material parameters of individual phases~\cite{Hexcel,Ohlhorst:CCdata}}
\label{T:phases}
\bigskip
\centering
\begin{tabular}{|c|c|c|c|}
\hline
Material &  Thermal conductivity & Specific heat & Mass density\\
& [Wm$^{-1}$K$^{-1}$] & [Jkg$^{-1}$K$^{-1}$] & [kgm$^{-3}$]\\
\hline 
Carbon fibers & (0.35, 0.35, 35) & 753   & 1810\\
Carbon matrix                 & 6.3      & 1256  & 1400\\
Voids filled with air         & 0.02     & 1000  & 1.3\\
\hline
\end{tabular}
\end{table}

For illustration, let us now consider an eight-layer carbon-carbon
composite laminate. Individual plies are made of plain weave carbon
fabric Hexcel 1/1 embedded in a carbon matrix. Each filament (fiber
tow) contains about 6000 carbon fibers T800H based on Polyacrylonitril
precursor. Such fibers are known as having a relatively low
orderliness of graphen planes on nano-scale. Nevertheless, unlike
glassy carbon, they still posses a transverse isotropy with the value
of longitudinal thermal conductivity considerably exceeding the one in
the transverse direction, see Table~\ref{T:phases}. As already
mentioned in the introductory part the carbon matrix forms as a result
of several cycles of carbonization/densification and final
graphitization of the initial polymeric precursor (green composite)
where the phenolic resin UMAFORM LE is used as the bonding agent
(matrix). Note that phenolic resins belong to a family of
non-graphitizing resins so that the final carbon matrix essentially
complies, at least in terms of its structure, with the original
cross-linked polymeric precursor. Therefore, the resulting material
symmetry is more or less isotropic with material parameters
corresponding those of glassy
carbon~\cite{Savage:1993:CCC,Fitzer:1998:CCC}.
\begin{figure}
\begin{center}
\includegraphics[width=12cm]{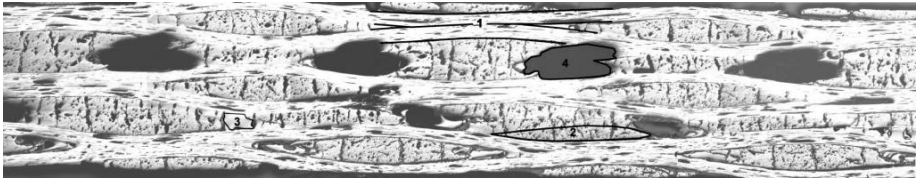}
\caption{Representative segment of eight-layer plain weave fabric laminate.}
\label{F2}
\end{center}
\end{figure}

A typical segment of the composite laminate appears in Fig.~\ref{F2}
showing characteristic porosity which may exceed at the structural
level (macroscale) 30$\%$~\cite{Tomkova:2004a,Tomkova:2004b} and is
often considered as an intrinsic property of this type of
composite~\cite{Savage:1993:CCC}. Several such micrographs were
processed with the help of LUCIA G~\cite{LIM}, Adobe Photoshop, Corel
Draw and Matlab R12 softwares to acquire information regarding the
basic structural units like an average thickness of carbon tows, size
of voids, shape and essential dimensions of fiber tow cross-section,
distribution of transverse and delamination cracks etc. which were
subsequently exploited in the construction of representative unit
cells on individual scales.

\subsection{Micro-scale}\label{sec:micro}
Starting with the fiber tow composite as the basic structural element
we recall Fig.~\ref{F1} showing a typical shape of the fiber tow
cross-section and significant amount of transverse cracks and voids
resulting in a non-negligible porosity up to 15\%. Unfortunately, a
detailed analysis of the fiber tow cross-section would be
computationally infeasible. As a suitable method of attack appears on
the other hand formulation of a two-step homogenization problem.

\begin{figure}[ht]
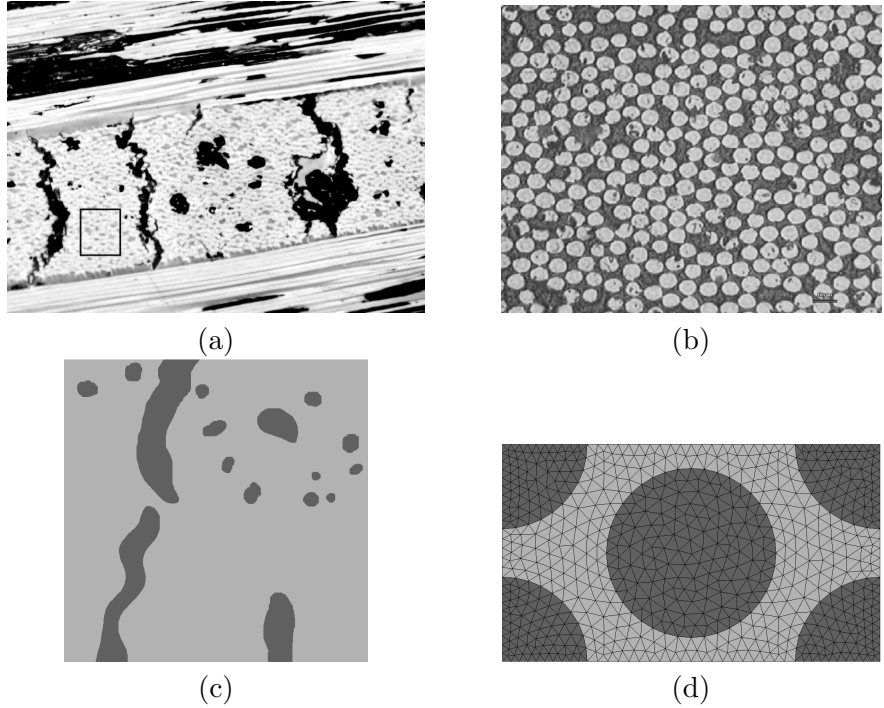

\begin{center}
\begin{tabular}{c@{\hspace{10mm}}c}
\includegraphics[width=5.5cm]{\figName{figure3a}} &
\includegraphics[width=5.0cm]{\figName{figure3b}} \\
(a)&(b)\\
\includegraphics[height=4cm]{\figName{figure3c}} &
\includegraphics[width=5cm]{\figName{figure3d}}\\
(c)&(d)
\end{tabular}
\caption{Homogenization on micro-scale: (a) fiber tow composite, (b)
  fiber-matrix composite, (c) voids-composite periodic unit cell, (d)
  fiber-matrix periodic unit cell}
\label{F3}
\end{center}
\end{figure}

To proceed, let us consider a typical micrograph of the fiber matrix
composite shown in Fig~\ref{F3}(b) taken as a random cut from the
fiber tow cross-section evident in Fig.~\ref{F3}(a). Providing this
section is sufficiently large to be statistically representative of a
real microstructure it becomes possible to proceed in the footsteps of
our previous work~\cite{Zeman:2001:EPG} and formulate a statistically
equivalent periodic unit cell at the level of individual
fibers. Keeping in mind on the other hand a relatively high volume
fraction of fibers, approx. 55\%, the results presented
in~\cite{Zeman:2001:EPG} and remarks put forward
in~\cite{Teply:1988:HEX} allows us to conclude that the actual
microstructure can be replaced with a simple periodic hexagonal unit
cell plotted in Fig.~\ref{F3}(d). The resulting effective material
parameters then serve as direct input for the analysis at the level of
the fiber tow cross-section. Here, a new unit cell, again exploiting
information acquired from image analysis, is introduced to properly
account for the porous phase. The proposed periodic unit cell, which
not only reflects the voids volume fraction but to some extent also
their arrangement, is seen in Fig.~\ref{F3}(c).

\subsection{Meso-scale}\label{sec:meso}
\begin{figure}[ht]
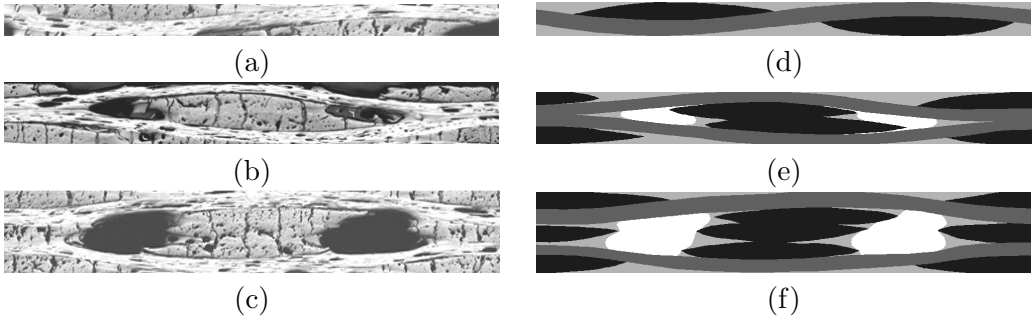

\begin{center}
\begin{tabular}{c@{\hspace{5mm}}c}
\includegraphics[width=6.5cm]{\figName{figure4a}}&
\includegraphics[width=6.5cm]{\figName{figure4b}}\\
(a)&(d)\\
\includegraphics[width=6.5cm]{\figName{figure4c}}&
\includegraphics[width=6.5cm]{\figName{figure4d}}\\
(b)&(e)\\
\includegraphics[width=6.5cm]{\figName{figure4e}}&
\includegraphics[width=6.5cm]{\figName{figure4f}}\\
(c)&(f)
\end{tabular}
\caption{Homogenization on meso-scale: 
(a)--(b) PUC1 representing carbon tow-carbon matrix composite, 
(c)--(d) PUC2 with vacuoles aligned with delamination cracks due to slip of textile plies, 
(e)--(f) PUC3 with extensive vacuoles representing the parts with textile reinforcement reduction due to bridging effect in the middle ply}
\label{F4}
\end{center}
\end{figure}

Having derived the effective material parameters for the fiber tow
composite, the assumed three level homogenization procedure continues
along the same lines on meso-scale. To that end, let us recall the
representative section of the composite laminate in Fig.~\ref{F2}.  A
detailed inspection of this micrograph reveals three more or less
periodically repeated geometries. For better view we refer to
Figs.~\ref{F4}(a)--(c).

Several such sections taken from various locations of the laminated
plates were examined, again with the help of image analyzer LUCIA G,
to obtain averages of various parameters including segment dimensions,
fiber tow thickness, shape of the fiber tow cross-section also
position, size and location of large vacuoles. Approximately 100
measurements were carried out for each segment and subsequently
utilized in the formulation of corresponding periodic unit cells
displayed in Figs.~\ref{F4}(d)--(f).

\subsection{Macro-scale}\label{sec:macro}
The final, clearly the most simple, step requires a construction of the
homogeneous, thought not isotropic, laminated plate. The stacking
sequence of individual periodic unit cells, which were introduced in
the previous section, complies with that observed for the actual
composite sample. Partially for the sake of simplicity, but also to be
consistent with the analyses performed on lower scales, the periodic
boundary conditions are considered even on the macro-scale. Providing
we are interested only in the bulk response of the laminate thus
ignoring detailed variation of the local fields, this assumption does
not yield a significant error in the desired estimates of the
macroscopic coefficients of thermal conductivities. Clear evidence is
available in Section~\ref{sec:res} comparing the numerical and
experimental results where the former ones are derived from the
theoretical grounds presented in the next section.
 
\section{Theoretical formulation}\label{sec:theory}
We now proceed to establish a framework for the determination of the
effective thermal conductivities regardless of the material system
considered in the previous section providing the same boundary
conditions are applied on each scale. An attentive reader will notice
a close similarity with the derivation of the effective elastic
material constants, see e.g.~\cite{Michel:1999:EPC} for that matter.

\subsection{Governing equations}\label{sec:gover}
Adhering to indicial notation with $\dot{a} =
\displaystyle{\frac{\partial a}{\partial t}}$ and $a_{,i} =
\displaystyle{\frac{\partial a}{\partial x_i}}$ representing the time
and space derivatives, respectively, the simplest form of the balance
equation reads
\begin{equation}
(\rho^s{C}_p^s)\dot{\theta}+q^s_{i,i} = 0,\label{eq:balance}
\end{equation}
where $\rho^s$ is the mass density and $C_p^s$ represents the specific
heat of a given phase $s$. If referring to micro-scale, for example,
the superscript $s$ may first represent the fiber or the matrix phase
and in the second step of homogenization, recall
Section~\ref{sec:micro}, it may be associated with the fiber-matrix
composite and voids. The phase constitutive equations follow from the
generalized version of Fourier's law and are provided by
\begin{equation}
q^s_i = - \chi_{ij}^sh^s_i,\hspace{0.75cm} h^s_i = -\psi_{ij}^sq^s_i,\hspace{0.75cm}\ins\Omega^s,\label{eq:fourier}
\end{equation}
where $h_i=\theta_{,i}$, $\tenss{\chi}$ is the thermal conductivity matrix
[Wm$^{-1}$K$^{-1}$] and $\tenss{\psi}=\tenss{\chi}^{-1}$ is the thermal
resistivity matrix. Providing a heat flux, consistent with
Eq.~\eqref{eq:THbc-1}$_2$, is imposed over the entire boundary
$\Gamma$ we arrive, with the help of Eq.~\eqref{eq:fourier}$_1$, at
the following boundary condition
\begin{equation}
\overline{q}_n = - n_i\chi_{ij}\theta_{,j}\hspace{1cm}\on\Gamma.\label{eq:fourier_BC}
\end{equation}
The weak form of Eqs.~\eqref{eq:balance}--~\eqref{eq:fourier_BC} is
then given by
\begin{equation}
\int_\Omega\left[\delta\theta(\rho{C}^p)\dot\theta+\delta\theta_{,i}\chi_{ij}\theta_{,j}\right]\de\Omega
+ \int_\Gamma\delta\theta\overline{q}_n\de\Gamma=0.\label{eq:PVW}
\end{equation}
It is interesting to show that under the steady-state conditions
($\dot\theta = 0$). Eq.~\eqref{eq:PVW} is essentially equivalent to
\begin{equation}
\avgs{\delta\theta_{,i}\chi_{ij}\theta_{,j}}=-\delta{H}_iQ_i,\label{eq:Hill}
\end{equation}
where $\avgs{a}$ represents a volume average of a given quantity,
i.e. $\avgs{a} = \frac{1}{|\Omega|}\int_\Omega a \de \Omega$.
This becomes evident after introducing Eqs.~\eqref{eq:Tloc-1}
and~\eqref{eq:THbc-1}$_2$ into the second term of Eq.~\eqref{eq:PVW}
to get
\begin{eqnarray}
\frac{1}{|\Omega|}\int_\Gamma\delta(H_ix_i+\theta^{*})Q_jn_j\de\Gamma
&=& \frac{1}{|\Omega|}\left[\delta{H}_iQ_j\int_\Gamma{x}_in_j\de\Gamma +
\int_\Gamma\delta\theta^{*}Q_jn_j\de\Gamma\right]\nonumber\\ &=&
\delta{H}_iQ_j\delta_{ij} = \delta{H}_iQ_i.
\end{eqnarray}
The integral $\int_\Gamma\delta\theta^{*}Q_jn_j\de\Gamma$ disappears
providing we set the fluctuation part of the temperature field
$\theta^{*}$ equal to zero on $\Gamma$ or impose the periodic boundary
conditions (the same values of $\theta^{*}$ on opposite sides of a
rectangular periodic unit cell). Either choice of variation of $\theta^*$ on $\Gamma$ readily ensures that $\avgs{h_i}=H_i$ since
\begin{equation}
H_i=\frac{1}{|\Omega|}\int_\Omega h_i\de\Omega = H_i+\frac{1}{|\Omega|}\underbrace{\int_\Gamma\theta^{*}n_i\de\Gamma}_{=0},\label{eq:volume_h}
\end{equation}
where integration by parts was used to transform the volume integral
into the integral over the boundary $\Gamma$. As already mentioned in
the introductory part, the periodic boundary conditions will be used
in the actual numerical analysis.

Note that Eq.~\eqref{eq:Hill} essentially resembles the Hill lemma in
the context of pure thermo-mechanical
problem~\cite{Sejnoha:Habil}. Here it shows the consistency of the
entropy change at the two associated scales (micro-meso,
meso-macro)~\cite{Auriault:1983:EMD,Ozdemir:IJNME:2008}.

The boundary conditions~\eqref{eq:THbc-1}$_2$ further imply the
volume average of the local heat flux be equal to the prescribed
macroscopic heat flux $Q_i$
\begin{equation}
\avgs{q_i} = Q_i.
\end{equation} 
This immediately follows under steady state conditions since $q_{i,i}
= 0$ so that $q_i = (q_jx_i)_{,j}$.  The volume average of the local
quantity $q_i$ then gives, see also~\cite{Ozdemir:IJNME:2008},
\begin{eqnarray}
|\Omega|\avgs{q_i} &=& \int_\Omega q_i\de\Omega\;=\;\int_\Omega(q_jx_i)_{,j}\de\Omega\nonumber\\
&=&\int_\Gamma q_jn_jx_i\de\Gamma\;=\;\int_\Gamma q_nx_i\de\Gamma\;=\;\int_\Gamma Q_jn_jx_i\de\Gamma\nonumber\\
&=&Q_j\int_\Omega x_{i,j}\de\Omega\;=\;|\Omega| Q_j\delta_{ij}\;=\;|\Omega| Q_i.\label{eq:volume_q}
\end{eqnarray}

\subsection{Effective conductivity and resistivity matrices}\label{sec:effective}
To proceed, we limit our attention to steady state conditions and
employ, in view of the forthcoming finite element formulation, the
standard matrix notation; e.g.~\cite{Bittnar:1996:NMM}. Then, under
pure thermal loading consistent with the boundary
conditions~\eqref{eq:THbc-1}$_1$ and taking into account the fact that
variation of a prescribed quantity vanishes, we receive the following
form of Eq.~\eqref{eq:Hill}
\begin{equation}
\avgs{\delta\evek{h}\trn\emtrx{\chi}\evek{h}}=0,\label{eq:Hill-E}
\end{equation}
In the framework of the finite element method (FEM) the vector
$\evek{h}$, recall Eq.~\eqref{eq:Tloc-1}, is provided by
\begin{equation}
\evek{h}=\evek{H}+\emtrx{B}\evek{\theta_d^{*}},\label{eq:h-FEM}
\end{equation}
where $\emtrx{B}$ stores the derivatives of the shape functions and
$\evek{\theta_d^{*}}$ lists the nodal values of the fluctuation part
of the temperature field. Substituting from Eq.~\eqref{eq:h-FEM} back into
Eq.~\eqref{eq:Hill-E} yields the resulting system of equations
\begin{equation}
\emtrx{K}\evek{\theta_d^{*}} = \evek{R},\label{eq:GE}
\end{equation}
where
\begin{eqnarray}
\emtrx{K} &=& \int_\Omega\emtrx{B}\trn\emtrx{\chi}\emtrx{B}\de\Omega,\label{eq:K}\\
\evek{R} &=& -\int_\Omega\emtrx{B}\trn\emtrx{\chi}\evek{H}\de\Omega\,=\,-\emtrx{S}\trn\evek{H}\label{eq:R}.
\end{eqnarray}
Solving for $\evek{\theta_d^{*}}$ from Eq.~\eqref{eq:GE} gives
\begin{equation}
\evek{\theta_d^{*}} = - \emtrx{K}^{-1}\emtrx{S}\trn\evek{H}\,=\,-\emtrx{G}\evek{H}.
\end{equation}
Next, introducing the vector $\evek{\theta_d^{*}}$ back into
Eq.~\eqref{eq:h-FEM} provides the volume average of the local heat
flux in the form
\begin{equation}
\evek{Q} = \avgs{\evek{q}} = -\frac{1}{|\Omega|}\int_\Omega\emtrx{\chi}\big[\emtrx{I}-\emtrx{B}\emtrx{G}\big]\de\Omega\evek{H}.
\end{equation}
Finally, writing the macroscopic constitutive law in the form
\begin{equation}
\evek{Q}=-\emtrx{\chi}^{hom}\evek{H},\label{eq:avg_h}
\end{equation}
readily provides the homogenized effective conductivity matrix $\emtrx{\chi}^{hom}$ as
\begin{equation}
\emtrx{\chi}^{hom} = \frac{1}{|\Omega|}\int_\Omega\emtrx{\chi}\big[\emtrx{I}-\emtrx{B}\emtrx{G}\big]\de\Omega.\label{eq:xi_hom}
\end{equation}

In actual computations the coefficients of the effective conductivity
matrix $\emtrx{\chi}^{hom}$ are found as volume averages of the local
fields from the solution of three successive steady state heat
conduction problems. To that end, the periodic unit cell is loaded, in
turn, by each of the two (2D) or three (3D) components of $\evek{H}$,
while the others vanish. The volume flux averages normalized with
respect to $\evek{H}$ then furnish individual columns of
$\emtrx{\chi}^{hom}$. The required periodicity conditions (the same
temperatures $\evek{\theta_d^{*}}$ on opposite sides of the unit cell)
are accounted for through multi-point constraints. In our particular
case it suffice to assign the same code numbers to respective periodic
pairs.

The derivation of the effective resistivity matrix may proceed along
the same lines providing the unit cell is loaded by the prescribed
macroscopic uniform heat flux $\evek{Q}$. In this particular case, the
volume average of the local temperature gradient is not known a
priori. Eq.~\eqref{eq:Hill} then yields two sets of governing
equations for unknown nodal values of $\theta^{*}$ and volume average
of the local temperature gradient $\avgs{\evek{h}}=\evek{H}$ in the
form
\begin{equation}
\emtrx{
\begin{array}{cc}
\emtrx{L} & \emtrx{S}\\
\emtrx{S}\trn & \emtrx{K}
\end{array}
}
\left\{
\begin{array}{c}
\evek{H}\\
\evek{\theta_d^{*}}
\end{array}
\right\}
=
\left\{
\begin{array}{c}
-\evek{Q}\\
\evek{0}
\end{array}
\right\},
\end{equation}
where the matrices $\emtrx{S}$ and $\emtrx{K}$ were already introduced
by Eqs.~\eqref{eq:K} and~\eqref{eq:R} and the matrix $\emtrx{L}$ is given by
\begin{equation}
\emtrx{L} = \int_\Omega\emtrx{\chi}\de\Omega.
\end{equation}
Combining the above equation together with the macroscopic constitutive law
\begin{equation}
\evek{H} = -\emtrx{\psi}^{hom}\evek{Q},
\end{equation}
gives after some manipulations the searched homogenized effective
resistivity matrix $\emtrx{\psi}^{hom}$ in the form
\begin{equation}
\emtrx{\psi}^{hom} = \big[\emtrx{L}-\emtrx{S}\emtrx{K}^{-1}\emtrx{S}\trn\big]^{-1}.
\end{equation} 

\section{Results of the numerical analysis}\label{sec:res}
The present section summarizes the numerical results of the proposed
three step uncoupled homogenization scheme. On each scale the relevant
periodic unit cell developed in Section~\ref{sec:RVE} was discretized
into finite elements. The geometrical complexity of individual
computational models together with the desired periodicity constraints
led to extremely fine meshes. While this may seem irrelevant from the
steady state conditions point of view, it suddenly becomes very
important when running the transient heat conduction problem.  This is
also why the periodic boundary conditions were disregarded in the
latter case. While the results are again presented separately for
individual scales, it is worthwhile mentioning that the output in
terms of the effective conductivities derived at a lower scale served
directly as the input for the up-scale analysis.  Regardless of the
type of analysis performed, steady state or transient conduction
problem, only the boundary conditions of type~\eqref{eq:THbc-1}$_1$
were considered and the effective conductivities were found through
the procedure described in Section~\ref{sec:effective}.

\subsection{Micro-scale}\label{sec:micro_2}
\begin{figure}[ht]
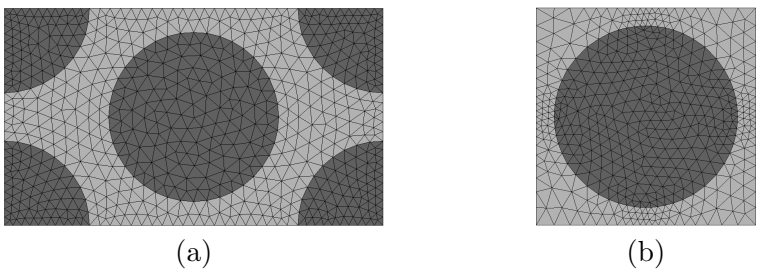

\begin{center}
\begin{tabular}{c@{\hspace{20mm}}c}
\includegraphics[width=5.0cm]{\figName{figure5a}}&
\includegraphics[width=2.9cm]{\figName{figure5b}}\\
(a)&(b)
\end{tabular}
\caption{Finite element mesh of fiber tow: (a) hexagonal arrangement of fibers, (b) square arrangement of fibers}
\label{F5}
\end{center}
\end{figure}

As suggested in Section~\ref{sec:micro} the effective conductivities of
the fiber tow composite were derived in two steps. First, an intact
carbon fiber-carbon matrix composite was considered. Two computational
models displayed in Fig.~\ref{F5} were examined. Both 2D and 3D
analysis was carried out. A simple rule of mixture (RM) was used to
estimate the effective thermal conductivities in the fiber direction
($z$-direction) for comparison with the general 3D analysis
\begin{equation}
k_z = c^fk_z^f+c^mk_z^m,
\end{equation}
where $c^f, c^m, k_z^f, k_z^m$ represent the fiber and matrix volume
fractions and conductivities in the $z$-direction, respectively. Their
values are listed in Table~\ref{T:phases}. The homogenized
conductivities then appear in Table~\ref{T:tow}.

\begin{table}[ht]
\caption{Effective thermal conductivities [Wm$^{-1}$K$^{-1}$] -  Intact fiber tow}
\label{T:tow}
\bigskip
\centering
\begin{tabular}{|c|c|c|c|c|}
\hline
Geometry & $k_x$ & $k_y$ & $k_z$ - 3DFEM & $k_z$ - RM\\
\hline 
hexagonal & 2.17 & 2.17 & 21.93 & 21.96\\
square    & 2.10 & 2.10 & 22.04 & 21.96\\
\hline
\end{tabular}
\end{table}

\begin{figure}[ht]
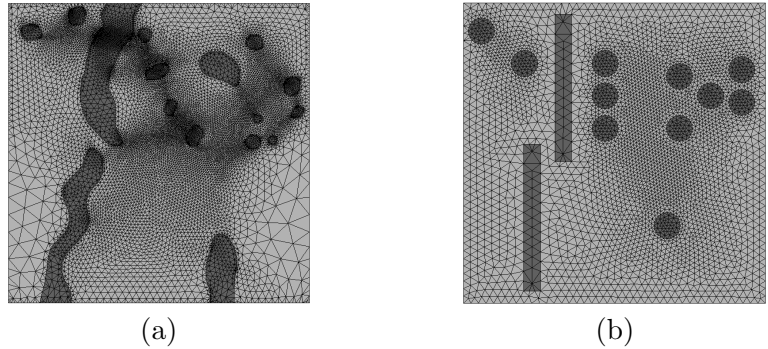

\begin{center}
\begin{tabular}{c@{\hspace{20mm}}c}
\includegraphics[width=4cm]{\figName{figure6a}}&
\includegraphics[width=4cm]{\figName{figure6b}}\\
(a)&(b)
\end{tabular}
\caption{Finite element mesh of fiber tow composite including voids: (a) real distribution of voids, (b) approximate distribution of voids}
\label{F6}
\end{center}
\end{figure}

As expected, there is a minor difference in the results provided by
both models. Nevertheless, the effective values derived from the
hexagonal array model were further employed in the subsequent analysis
step which allowed us to introduce the intra-tow voids into the intact
but already homogeneous fiber tow composite. Again, two computational
models evident from Fig.~\ref{F6} were studied. 

Hereafter, we were concerned with the two-dimensional problem only.
The corresponding results are available in Table~\ref{T:voids}.  Apart
from a steady state analysis a transient heat conduction problem
goverened by Eq.~\eqref{eq:balance} was addressed. Clearly, solving
this problem then calls for the effective mass density and specific
heat on the level of an intact fiber/matrix composite. For this
purpose a logarithmic rule of mixture was adopted in the form
\begin{eqnarray}
\ln\rho &=& c^f\ln\rho^f + c^m\ln\rho^m,\\
\ln C_p &=& c^f\ln C_p^f + c^m\ln C_p^m.
\end{eqnarray}
The analysis was performed with the help of FEMlab commercial code
using the Heat Transfer Mode~\cite{FEMlab}. To simulate a
unidirectional flow the unit cell was heated on the one side while
zero temperature was prescribed on the other side. The initial
temperature was also assumed to be equal to zero. In addition, a zero
flux boundary conditions were prescribed on the remaining sides to
represent isolated surfaces. This clearly yields the mixed boundary
conditions on $\Gamma$, which in turn may violate the
conditions~\eqref{eq:volume_h} and~\eqref{eq:volume_q} if the periodic
boundary conditions are disregarded, see the discussion in the
following paragraphs. The conservation condition~\eqref{eq:Hill},
however, holds even in this case. 

The thermal conductivities were then estimated from
Eq.~\eqref{eq:avg_h} after reaching the steady state conditions (time
independent temperature profile). In such a case Eq.~\ref{eq:balance}
reduces to $q_{i,i}=0$ thus naturally enforcing
Eq.~\eqref{eq:volume_q}. If for example a unidirectional heat flow
along the macroscopic $x$-axis is considered, then the component of
the effective heat conductivity is, in view of Eq.~\eqref{eq:avg_h},
provided by
\begin{equation}
{\chi_x}^{hom} = -\frac{\langle{q_x}\rangle}{\langle{\theta}_{,x}\rangle}.\label{eq:chi-hom-gx}
\end{equation}
The ${\chi_y}^{hom}$ component of the effective conductivity matrix $\emtrx{\chi}^{hom}$ is
derived analogously.  If no action is taken this result corresponds to
the assumption of $t^*=0$ on $\Gamma$.

\begin{table}[ht]
\caption{Effective thermal conductivities [Wm$^{-1}$K$^{-1}$] -  Fiber tow with hexagonal arrangement of fibers including voids}
\label{T:voids}
\bigskip
\centering
\begin{tabular}{|c|c|c|c|c|c|c|}
\hline
Analysis & \multicolumn{3}{|c|}{Steady state} & \multicolumn{3}{|c|}{Transient}\\
\hline
Voids distribution & $k_x$ & $k_y$ & $k_z$ - RM & $k_x$ & $k_y$ & $k_z$ - RM\\
\hline 
real         & 0.97 & 1.70 & 19.01 & 1.20 & 1.60 & 19.10\\
approximate  & 1.12 & 1.77 & 19.01 & - &  - & - \\
\hline
\end{tabular}
\end{table}

The results in Table~\ref{T:voids} suggest a minor difference between
the steady state and transient heat analysis. Among others the lack of
periodic boundary conditions, much coarser mesh~\cite{Tomkova:2006}
and truly approximate estimates of the homogenized mass density and
specific heat appear as the most critical sources for the resulting
difference. One may also expect this difference to grow when moving up
the scales particularly if assuming a certain consistency of the three
step procedure in the sense that the results derived on a lower scale
using one type of analysis are transferred to a higher scale where the
same type of analysis is performed.

\subsection{Meso-scale}\label{sec:meso_2}

\begin{figure}[ht]
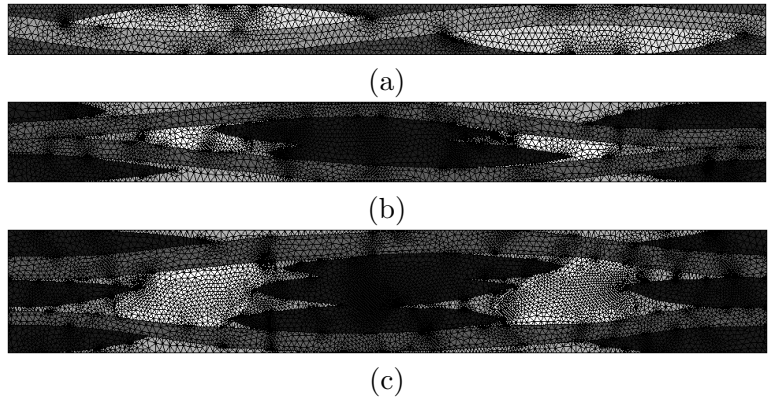

\begin{center}
\begin{tabular}{c}
\includegraphics[width=10cm]{\figName{figure7a}}\\
(a)\\
\includegraphics[width=10cm]{\figName{figure7b}}\\
(b)\\
\includegraphics[width=10cm]{\figName{figure7c}}\\
(c)
\end{tabular}
\caption{Finite element meshes; (a) PUC1, (b) PUC2, (c) PUC3.}
\label{F7}
\end{center}
\end{figure}

At this level the carbon fiber tow is treated as a homogeneous phase
with the effective material parameters derived from the numerical
analysis on micro-scale. Since these are provide in the local (fiber)
coordinate system, it is desirable, in order to account for the tow
waviness, to transform the corresponding effective conductivity matrix
into the global coordinate system as
\begin{equation}
\emtrx{\chi(\tenss{x})}^{tow}_g =
\emtrx{T(\tenss{x})}\trn\emtrx{\chi}^{tow}_l( \vek{x} )\emtrx{T(\tenss{x})},
\end{equation}
where $\emtrx{T(\tenss{x})}$ is the continuously varying
transformation matrix. The matrix $\emtrx{\chi(\tenss{x})}^{tow}_g$ then
enters Eq.~\eqref{eq:xi_hom} where applicable. The computational
procedure is, nevertheless, identical to the one outlined in the
previous section. With reference to Section~\ref{sec:meso} the three
periodic unit cells in Fig.~\ref{F7} were examined. Table~\ref{T:cell}
then provides the resulting homogenized thermal conductivities. To
accept the notable difference in the results from the two types of
distinct analyses we address the reader to the comments offered in the
last paragraph of Section~\ref{sec:micro_2}.

\begin{table}[ht]
\caption{Effective thermal conductivities [Wm$^{-1}$K$^{-1}$] -  Representative unit cells of textile composites}
\label{T:cell}
\bigskip
\centering
\begin{tabular}{|c|c|c|c|c|}
\hline
Analysis & \multicolumn{2}{|c|}{Steady state} & \multicolumn{2}{|c|}{Transient}\\
\hline
Cell & $k$-longitudinal & $k$-transverse & $k$-longitudinal & $k$-transverse \\
\hline 
PUC 1  & 9.46 & 2.27 & 9.10 & 2.30\\
PUC 2  & 9.03 & 1.47 & 7.40 & 1.80\\
PUC 3  & 7.29 & 1.53 & 6.60 & 1.75\\
\hline
\end{tabular}
\end{table}

\begin{figure}
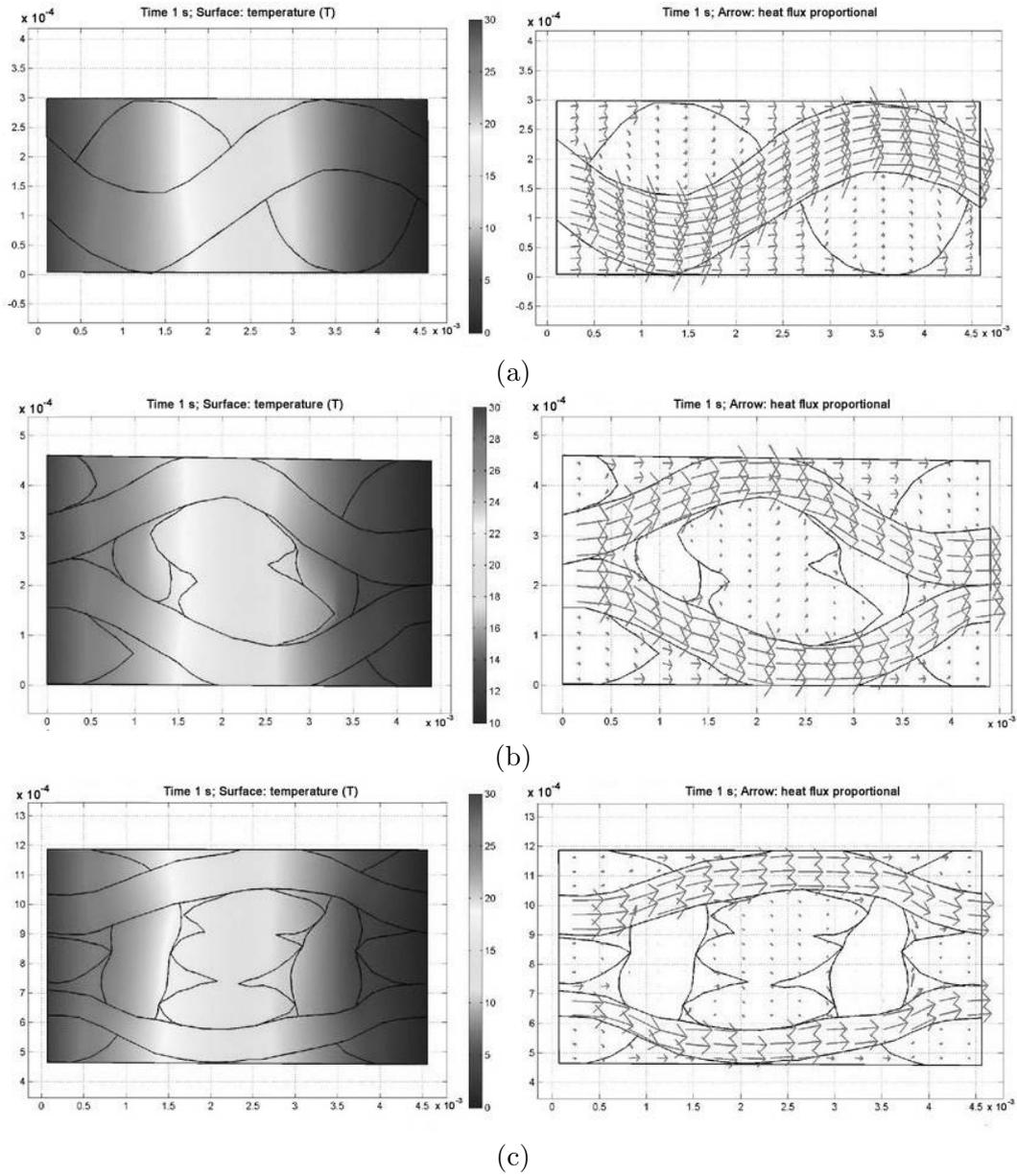

\begin{center}
\begin{tabular}{c}
\includegraphics[width=14cm]{\figName{figure8a}}\\
(a)\\
\includegraphics[width=14cm]{\figName{figure8b}}\\
(b)\\
\includegraphics[width=14cm]{\figName{figure8c}}\\
(c)
\end{tabular}
\caption{Simulation of transient heat conduction problem - temperature profile and heat flux 
in the direction parallel to the composite plate: (a) PUC1, (b) PUC2, (c) PUC3}
\label{F8}
\end{center}
\end{figure}

Although more complicated, solving the transient heat conduction
problem allows us to receive further information regarding the
response of a composite to thermal loading including a gradual
evolution of the temperature profile and time lag to reach the steady
state conditions. Fig.~\ref{F8} shows a typical graphical output of
the results provided by FEMlab. These plots, in particular, correspond
to the onset of steady state conditions showing already more or less
constant temperature gradient (linear variation of temperature) for
all unit cells with corresponding distribution of heat fluxes.

\subsection{Macro-scale}\label{sec:macro_2}
The macroscopic analysis of heat conduction problem represents the
final step of the proposed multiscale approach. From the computational
point of view it requires a relatively simple analysis of a
three-layer laminate stacked from three periodic unit cells shown
Fig.~\ref{F7}. However, the geometrical details of these unit cells
are no longer relevant. Instead, they are treated as homogeneous
blocks with the assigned homogenized (mesoscopic) properties. The
resulting macroscopic properties are stored in Table~\ref{T:comp}.

\begin{figure}
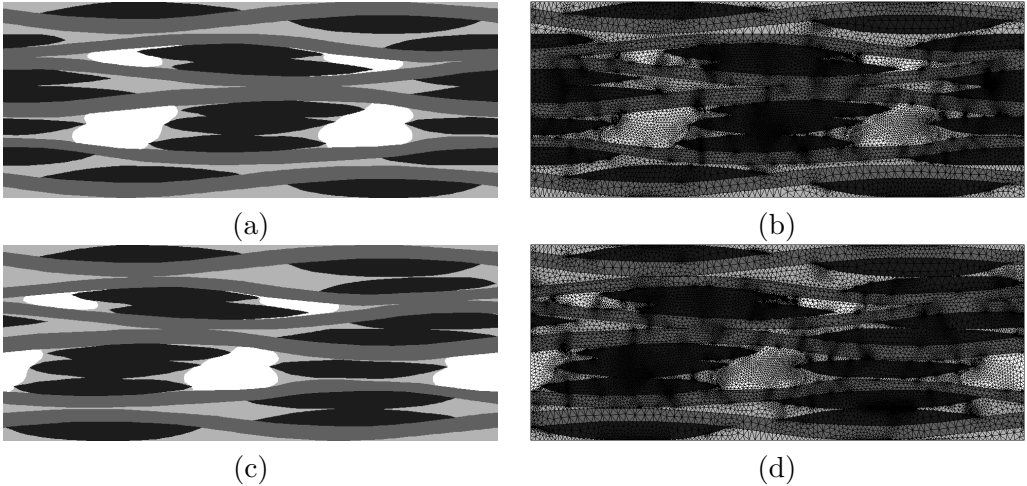

\begin{center}
\begin{tabular}{cc}
\includegraphics[width=6.5cm]{\figName{figure9a}}&
\includegraphics[width=6.5cm]{\figName{figure9b}}\\
(a)&(b)\\
\includegraphics[width=6.5cm]{\figName{figure9c}}&
\includegraphics[width=6.5cm]{\figName{figure9d}}\\
(c)&(d)
\end{tabular}
\caption{Textile laminate with: (a)--(b) regular arrangement of plies,
  (c)--(d) irregular arrangement of plies.}
\label{F9}
\end{center}
\end{figure}

To validate these results an additional numerical analysis was
performed. Here, all geometrical details of the laminated plate were
taken into account by creating a multi-layered periodic unit cell
exploiting the geometrical models developed for the mesoscopic
analysis. The selected stacking sequence was motivated by images of
the real laminate such as the one displayed in Fig.~\ref{F2}. Two such
unit cells were constructed, see Fig.~\ref{F9}. While the first unit
cell, Fig.~\ref{F9}(a), assumed an ideal stacking with regular
arrangement of individual unit cells, the second one,
Fig.~\ref{F9}(c), allowed for a mutual shift of individual plies to
approximate the actual geometry more accurately. It is interesting to
see, Table~\ref{T:comp}, that all types of analyses (laminate, PUC
with regular and PUC with irregular arrangement of unit cells)
essentially provide the same estimates of the macroscopic thermal
conductivities.  Such a conclusion clearly advocates the use of
multiscale analysis at least in the present context of linear steady
state or transient heat conduction problem.

\begin{table}[ht]
\caption{Effective thermal conductivities [Wm$^{-1}$K$^{-1}$] -
  Laminated plate~(The number in parentheses indicates the difference between a
  numerical value and experimental data.)}
\label{T:comp}
\bigskip
\centering
\begin{tabular}{|c|c|c|c|c|c|}
\hline
Analysis & \multicolumn{2}{|c|}{Steady state} & \multicolumn{2}{|c|}{Transient}\\
\hline
Geometry & $k$-longitudinal & $k$-transverse & $k$-longitudinal & $k$-transverse \\
\hline 
Laminate             & 8.47~(15.3$\%$) & 1.66~(3.75$\%$) & 8.30~(17.0$\%$) & 1.80~(12.5$\%$)\\
Macro cell regular   & 8.65~(13.5$\%$) & 1.68~(5.00$\%$) & -     & -   \\
Macro cell irregular & 8.60~(14.0$\%$) & 1.67~(4.37$\%$) & -     & -   \\
Measured             & -     & -       & 10.00 & 1.60\\
\hline
\end{tabular}
\end{table}

To judge the quality of any computational approach purely from
numerical experiments seems, however, rather shallow. To enhanced
credit of a numerical analysis it is therefore desirable to compare
the numerical results with those obtained experimentally. For this
type of composite the results from experimental investigation of the
thermophysical properties are available in~\cite{Kubicar:2002}. For
details on the pulse transient method used in this work together with
computational models required to relate temperature to the generated
heat pulse we refer the reader to the above paper. Some basic
information can also be found in~\cite{Tomkova:2006}. The
experimentally determined macroscopic conductivities appear in
Table~\ref{T:comp}. Taking into account the possible errors in the
determination of phase material parameters (carbon fibers and carbon
matrix) on the one hand and errors associated with the laboratory
measurements on the other hand adds further confidence in the
presented three-level uncoupled multiscale homogenization approach.
In this context, the simplified two-dimensional approach (with the
exception to the lowest scale) adopted for the solution of a generally
three-dimensional problem, a natural component of laboratory
measurements, should also be added to a list of sources of possible
errors. It is worth noting even under these simplifications, the
errors associated with the analysis compare well with the results of
fully 3D analyses reported in~\cite{Woo:2004:TCCP}. Therefore, it is
expected that a reliable three-dimensional geometrical model
constructed on the mesoscopic level will further improve the
predictive capabilities of the multiscale solution strategy. This is
the topic of our current research.

\section{Conclusions}\label{sec:con}
Three levels of hierarchy are introduced in this contribution to
derive the effective thermal conductivities of a plain weave textile
laminate. Different resolution of microstructural details are
considered on individual scales for the construction of an adequate
representative unit cell. Such a unit cell arises as a result of
elaborate evaluation of images of a real composite sample. The
geometrical complexity of these types of composites are mainly
responsible for a slow progress in the formulation of a generally
three-dimensional unit cell. Possibility to enlighten this subject is
given in~\cite{Zeman:2004:RC} promoting the construction of such an
RVE by matching statistical characteristics of both the real composite
and RVE, at this step, however, in the absence of a porous
phase. Owing to a significant contribution of this phase to the
overall volume of the composite ruled the choice of a simplified
two-dimensional analysis for the present study.

Clearly, the theoretical formulation, here developed on the basis of
the first-order homogenization, is quite general and space invariant.
Both transient and steady state conditions were examined in this
study. When considering solely the temperature boundary conditions of
type~\eqref{eq:THbc-1}$_1$ allows the solution to be accepted even if
neglecting the periodic boundary conditions. Proper averaging
relations then still provide the correct means for the scale
transition. This has been exploited when running the transient heat
conduction problem. The periodic constraints, on the other hand, were
imposed for the estimates of effective thermal conductivities under
the steady state conditions. In this case, too fine discretization
needed for the straightforward introduction of periodic boundary
conditions was of minor concern. Both approaches have shown, however,
their potential in the derivation of the desired effective
(homogenized) thermal conductivities of highly complex plain weave
textile composites through the application of fully uncoupled
multiscale homogenization scheme.

\section*{Acknowledgments}
The financial support provided by the GA\v{C}R grant No.~106/07/1244
and partially also by the research project CEZ~MSM~6840770003, is
gratefully acknowledged.



\end{document}